# Moving and resizing of the screen objects

The shape and size of the objects, which we see on the screen, when the application is running, are defined at the design time. By using some sort of adaptive interface, developers give users a chance to resize these objects or on rare occasion even change, but all these changes are predetermined by a developer; user can't go out of the designer's scenario. Making each and all elements moveable / resizable and giving users the full control of these processes, changes the whole idea of applications; programs become user-driven and significantly increase the effectiveness of users' work. This article is about the instrument to turn any screen object into moveable / resizable.

## Introduction

More than 20 years ago DOS gave place to multi windows operation systems. From that time each application has its own window, which can be moved and resized by users at any moment. Users make the decision about the importance of each application and about the screen space that each application can take. Each window gets the starting size and position either according with the designer's ideas or by default, but user can easily change these parameters at any moment. <u>On the upper level users decide what the best for their work is</u>.

Upper level is mostly informative level; the real work is done inside the applications: input of data, calculations, transformation of the results, and visualization are done only on the inner level. Surprisingly, users are stripped of all the chances to rule at the inner level. This level is much more important for the real work, but here users can only do whatever the designers allow them to; users have no chances to transform the inner view of the applications.

Well, they can do something, but this "something" is always inside the scenario, written by the designers. The fixed "inner world" is not good even if you are the only user of your own program, because even for a tiny change in the view you have to rewrite your code and recompile it. More common case, when one or several people design an application, used by hundreds, thousands or millions, and certainly not all of them share the designer's aesthetic views. It can be not so important for the simple applications, like *Calculator*, but there are much more complicated applications, like those used in engineering or scientific areas, and there is not a rare situation when the users are much more qualified and experienced than the designers, and yet the work must be done only inside the designers' vision. This greatly affects the users work, but this is the way the modern day applications are working.

The problem is well known and a lot of people work on softening this conflict. The good solution is often seen in the use of *adaptive interface*, which allows users to change the application's view according with their personal demands. There are tons of papers on this subject, and there are different approaches to build a good adaptive interface; usually it gives users an instrument to select among some set of the predefined cases. The number (variety) of cases and the easiness of selection depend on the programmer's level, but the main limitation and the weakest side of the adaptive interface that all these choices are predefined. One of the branches of adaptive interface is called *dynamic layout,* and Microsoft announced it as a main direction for the coming years. Under the implemented dynamic layout, there is no selection among the list of cases, but the sizes and positions of elements are determined based on the main window's size and the selected font's size. In the simple situations it can work perfectly; in more complicated situations it's nearly a disaster. It is certainly a wrong idea that the designer can give out something that will be the best for everyone and all on any computer configuration.

Is there any good solution? Yes, but in the orthogonal direction: let users decide, what the best is for each of them at any moment. Is there any hurdle on this way? Yes, a simple one: programmers have no easy instrument to write such kind of programs. Is it possible to design such instrument? Yes. Let's first look at the core of the problem, and then we'll look at the solution.

<u>Small remark</u>. I work under Windows system, use Visual Studio for design of my applications, and write them mostly in C#, so throughout this text I'll be using some familiar terms. But the ideas of this article are not limited to C# or mentioned operation system.

## Objects in applications

On the upper level there is one kind of elements – windows. Each window is resizable and moveable; these are their default features, and only for special reason they can be eliminated. On the inner level (inside any window) we have two kinds of elements: controls and graphical objects. In reality, whatever you see at the screen is a graphical object, but as was declared in a famous book [1] decades ago: "All animals are equal, but some animals are more equal than others". Controls are those "more equal" elements on the screen, and their behaviour is absolutely different from ordinary graphical objects.

*Controls* are moveable and resizable by default; Windows guarantee that all of them have these features. Though all controls are moveable and resizable, these features are not obvious and are used very rarely. Controls have no title bars, so there is no indication that they can be moved; usually there are no borders that indicate the possibility of resizing. But programmers can easily use these features for all controls, and from time to time they do use them, for example, via anchoring and docking. The most important thing is not *how* controls can be moved and resized, but that for them moving and resizing *can be organized* without problems. Controls have a lot of features, but for the purpose of this article two of them are very important:

- All the standard controls have rectangular shape.
- Controls can respond to mouse events, their reaction to mouse events is predetermined and expected by users, and it's better not to change it.

*Graphical objects* are of an absolutely different origin than controls and, by default, they are neither moveable, nor resizable. There are two well known techniques to make graphical objects look "resizable": the first one is to place such object on a panel and make the last one resizable, for example, via anchoring / docking; the second one is to use the bitmap operations. The use of these tricky solutions is very limited, both of them can be used only with the rectangular objects, and the bitmap operations are limited even more because of the quality problems.

The inner world of any application can be populated with graphical objects (ground) and controls (clouds). Objects on the ground can stay side by side or one atop of another, and clouds can be set aside or close each other from view, but any cloud is always above the ground. (Forget about couple of skyscrapers that can go above the clouds, such thing cannot happen in computer world.) Users can communicate with the controls, but only because Windows (and similar systems) allow us to do it. There is a special managing program (part of the system) that analyses every mouse click on the screen and sends the corresponding message to the appropriate control, if the click was inside the control's area. Controls are painted on the screen according with their hierarchy (Z-order), the highest control at the point receives the message, so this special program analyses the position of the controls and their hierarchy.

Mouse clicks outside the controls are not directed to any object, but can be analyzed by the program and reflected in some changes. Such clicks can be interpreted as direct commands to graphical objects, and in some rare cases it is done in this way. Some applications, like *Paint*, implemented such thing for several graphical classes, but for special classes anything can be done.

My goal was not the moving and resizing of some special objects (this I was doing for years), but the design of mechanism for any kind of objects. For the programmer, this mechanism must be easy in implementation, but very powerful, so that no object will be an exception that can't be involved in moving / resizing. For users, it must be an extremely easy technique of moving / resizing any element on the screen; users don't know anything about the difference of controls and graphical objects, they simply deal with whatever they see, so this mechanism must work identically with any kind of objects.

The designed mechanism of moving and resizing of the screen objects consists of two parts:

1. An object that is supervising all moving / resizing. This is an object of the `Mover` class.
2. Special additional feature for any screen object, which turns it into moveable / resizable. By analyzing this feature, `Mover` decides how this element must be moved and / or reconfigured. This feature is called *contour*.

Before going into further details of `Mover`'s work and contours' design, I want to emphasize several things.

- `Mover` is not working with the real objects, but only with their contours.
- `Mover` doesn't know about all the objects of a form (dialog); `Mover` works only with the contours that were included into its List<>.
- Contour is a special feature to be used by `Mover` and has no other effect.
- Contour makes an object moveable / resizable, but only if this object is registered with a `Mover`.
- The only instrument that is used to move / reconfigure objects is a mouse. If in the nearest years all the screens will become touchable, this will not influence the proposed technique. At the moment the mouse is like a glove on our hand: move the cursor (hand) to the needed point, press the button (grab an object or its part), move the grabbed part to the new location, and release it. Some objects need not only forward movement, but also rotation; to distinguish between these movements, left and right mouse buttons are used.

The further explanation is illustrated by the different samples from the working applications. The results of my work (in a form of several articles, working applications, and a full project with all the codes in C#) can be seen at (and downloaded from) www.sourceforge.net in the project **MoveableGraphics** (names of the projects are case sensitive there!); the new results are published there usually once a month.

Classes, used for explanation in this article, are from the **Test_MoveGraphLibrary** application (all codes are available there) and from **MoveGraphLibrary.dll**. If you start the **Test_MoveGraphLibrary.exe** and select the menu position *Abstract figures – Contour samples*, you'll open the **Form_ContourSamples.cs**, which demonstrates all the contours that are described further on.

## Contours and their design

Users don't know anything about the differences between controls and graphical objects. I want both types of elements to be involved in movement / resizing in identical way or nearly identical way, and this is going to be one of the challenges, because controls differ from others ("… are more equal then others"), and the proposed mechanism does not require any changes in the current operation system and works under Windows according with all its rules. Let's first deal with the graphical objects, and then apply the results to controls.

There are two standard ways to add moveable / resizable features to the objects: either to use an interface or an abstract class; after trying both ways I decided upon an abstract class. In my design, to make any graphical object moveable / resizable, it must be derived from the abstract class `GraphicalObject.` (There is an exception from this rule; you can read about it in the *Design and use of moveable and resizable graphics* in the chapter *Parallel way*.) The `GraphicalObject` class declares three crucial methods.

```
public abstract class GraphicalObject
{
    public abstract void DefineContour ();
    public abstract void Move (int cx, int cy);
    public abstract bool MoveContourPoint (int i, int cx, int cy, Point ptMouse,
                                           MouseButtons catcher);
    …
}
```

I want to implement both movement and resizing (or reconfiguring) of the objects. Moving / resizing must be done in the most natural way "grab – move – release", and as the only instrument to be involved is a mouse, then I see only one solution: to define some sensitive areas of an object, where the reconfiguring or resizing can be started, and other sensitive areas for moving of the whole object. At the beginning of the design, the areas for resizing were small, even tiny, and I called them *nodes*. The areas for moving objects were only around the lines that connected those nodes, so I called them *connections*. When I painted all those areas atop the picture of an object, those small spots, connected with the lines, looked like a *contour*, and that's where the word came from.

<p align="center">Contour = Nodes + Connections</p>

You'll see further on that with the new ideas in organizing those sensitive areas, their combinations often do not look like contours, but I continue to use this word. If a contour is applied to an object and this object is registered with a `Mover`, then moving / resizing is done regardless of whether the contour is shown or not; to indicate the contour under the mouse, the change of cursor is used.

**Nodes** (class `ContourApex`) are used as sensitive areas for reconfiguring or resizing. A movement of any node can be really free, or can be restricted by some (or many) limitations. A movement of some object's part can change the shape of an object (reconfiguring), or it can result in movement of other parts in such a way, as not to change the shape, but only the size of an object (resizing). Nodes may have different shapes and sizes. Each node is associated with some point; for square and circular nodes the location of this point is very important; for some polygon nodes it doesn't matter at all. The square and circle nodes are defined by the center point and the size; polygon nodes are defined by the location of vertices. These parameters define the node's sensitive area, and when the mouse cursor is moving across this area, the shape of the cursor can be changed to indicate that the node can be grabbed and moved. The node's area can be set to `null`; such type of nodes are used at the ends of connections, when only the moving of an object is needed, but not the resizing.

```
ContourApex (int nVal,              Identification number.
             Point ptReal,          A real point on the screen, associated with this node.
             Size szRealToSense,    Shift from the real point to the middle of the node.
             MovementFreedom mvt,   Possible sole movements of the node. {None, NS, WE, Any}
             Cursor cursorShape)    Cursor shape above the node.
```

Only the first two parameters are mandatory; others can be omitted in any combination or at all, and the default values are:
```
                           new Size (0, 0)
                           MovementFreedom .Any
                           Cursors .Hand
```

Each node is identified by its number; these numbers are later used as the parameters of the `MoveContourPoint()` method, which includes the code only for the nodes that can be moved individually. The shift of the node from the real point can be used to move the visualized contour aside from the image of the object. If this shift is not needed, then the third parameter can be omitted, and the default value will be set to zero. When some shift is needed, it can be defined as a separate parameter, or simply added to the definition of the real point. When the node's area is not null and the mouse cursor is moving across this area, the shape of the cursor is usually changed to indicate that the node can be grabbed and moved. There are some conformity rules for possible movements of the object and the shape the mouse cursor can become over it, but there is some flexibility in defining these pair of parameters.

The above shown version of constructor and its variants are used for square and circle nodes; for polygon nodes there is another basic form of constructor.

```
ContourApex (int nVal, Point ptReal,
             Point [] poly,
             Size szRealToSense, MovementFreedom mvt, Cursor cursorShape)
```

For polygon nodes the real point is often not important at all; it can be omitted and the last three parameters also. As polygon nodes are often used not for resizing / reconfiguring, but for moving of the whole object, the default value for the cursor above such nodes is changed to `Cursors.SizeAll`.

**Connections** (class `ContourConnection`) are used as sensitive areas for moving the whole object. Connection is a sensitive area around the straight line between the two nodes; the area is described by its sensitivity – maximum distance to the line, where it can be still grabbed.

```
ContourConnection (int A, int B,      Identification numbers of the connected nodes.
                   int sense)         Sensitivity of the connection.
```

The shape of the sensitive area is defined by the ratio between the length of the line and the width of the area: if the length is much bigger than the width, the area looks like a strip; if the width is much bigger than the length, it looks like a circle; the intermediate variants have a "sausage" shape. Usually the width of the sensitive areas around all the connections of the contour is the same; however, it is possible to set this width individually for each connection. When the mouse cursor is moving across the connection's area, the shape of the cursor is changed to indicate that the entire object can be grabbed and moved.

Classical contour consists of nodes and connections:

```
Contour (ContourApex [] ca, ContourConnection [] cc)
```

The contour constructor is overloaded; there are several basic variants, and later I'll demonstrate samples of contours, consisting only of nodes without any connections at all.

Movement of an object means relocation of some basic points, which define the position of the object, and repainting of this object at a new place. The sizes of an object are not changed during such movement, so the `Move()` method only describes the relocation of those basic points as the result of the mouse move.

The system of nodes and connections is a very flexible one. For any object, a contour can be designed in different ways, depending on the idea of sensitive areas; these areas are described in the `DefineContour()` method. Even for the same object, there can be different contours; each contour will require its own `MoveContourPoint()` method, as it is strictly linked with the movement of the particular nodes. On the other hand, the `Move()` method has nothing to do with the nodes or connections, except the fact that it is started by them.

The best way to understand the process of turning graphical objects into moveable / resizable, is to analyze together their contours definition and methods for moving these objects. The proposed samples are based on the simplest geometrical figures, which make these samples easy to understand, but my experience shows that even the most complicated screen objects have either identical contours or slightly modified.

**Case 1**. *Rectangle*.

   Moving by border; resizing by corners.

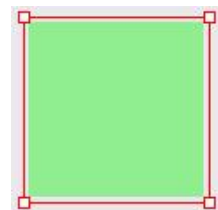

There will be several contours for moving and resizing of an ordinary rectangle area; in all these cases there is a limit on minimum rectangle size to avoid its disappearance.. The first rectangle (**figure 1**) can be moved by any border point; the resizing is done by moving any of the corners.

**Fig. 1**

```
public override void DefineContour ()
{
    int shift = 3;
    ContourApex [] ca = new ContourApex [4];
    ca [0] = new ContourApex (0, new Point (rc .Left – shift, rc .Top – shift));
```

```
    ca [1] = new ContourApex (1, new Point (rc .Right, rc .Top),
                              new Size (shift, -shift));
    ca [2] = new ContourApex (2, new Point (rc.Right + shift, rc .Bottom + shift));
    ca [3] = new ContourApex (3, new Point (rc .Left - shift, rc .Bottom + shift));
    ContourConnection [] cc = new ContourConnection [4] {
                new ContourConnection (0, 1), new ContourConnection (1, 2),
                new ContourConnection (2, 3), new ContourConnection (3, 0) };
    contour = new Contour (ca, cc);
}
```

I especially used slightly different form of constructor for one of the nodes to show that the shift from real point to the sensitive point can be either passed as a separate parameter, or combined with the point values.

Move(…) – the method of moving the whole object – is started by one of the contour's elements, but its code never mentions nodes or connections. On the contrary, this method only describes the movement of the real simple elements, on which any, even the most complicated objects, are built. In this case there is only one such element – rectangle.

```
public override void Move (int cx, int cy)
{
    rc .X += cx;
    rc .Y += cy;
}
```

MoveContourPoint(…) – the method of individual nodes movements. Such movements can be absolutely free, or there can be some restrictions for each involved node, or the nodes can influence each others relocation. It looks like with the increasing number of nodes this method will become really huge, but in reality it never happens. Further on there will be the cases with the hundreds of nodes, but as the behaviour of each node is identical to others, the code of the method is short enough.

Each of the four nodes in this case can change the width and height of the rectangle; each proposed resizing must be checked against the limitations, so the full code is a bit long to be shown here, but the individual movements of the nodes are partly identical, and the code is rather simple. Here is a code for the node in the right top corner.

```
public override bool MoveContourPoint (int i, int cx, int cy, …, MouseButtons btn)
{
    bool bRet = false;
    if (btn == MouseButtons .Left)
    {
        …
        else if (i == 1)     // Right-Top corner
        {
            if (rc .Height - cy >= minH)
            {
                rc .Y += cy;
                rc .Height -= cy;
                bRet = true;
            }
            if (rc .Width + cx >= minW) {
                rc .Width += cx;
                bRet = true;
            }
        }
        …
```

**Case 2**. *Rectangle*.
    Moving by border; resizing by corners.

Several differences (**figure 2**) from the previous case:
- Nodes are placed exactly in the corners of a rectangle without any shifts.
- The form of the nodes is changed from square to circle.
- The size of the nodes is slightly enlarged to make the resizing a bit easier to start.

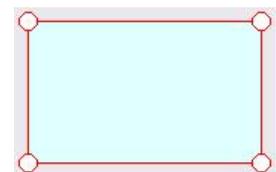

**Fig. 2**

These changes are reflected in the code of the DefineContour() method.

```
public override void DefineContour ()
{
    ContourApex [] nodes = new ContourApex [4] {
            new ContourApex (0, new Point (rc .Left, rc .Top)),
            new ContourApex (1, new Point (rc .Right, rc .Top)),
            new ContourApex (2, new Point (rc .Right, rc .Bottom)),
            new ContourApex (3, new Point (rc .Left, rc .Bottom)) };
    foreach (ContourApex ca in nodes)
    {
        ca .SenseAreaForm = NodeForm .Circle;
        ca .SenseAreaSize = 6;
    }
    contour = new Contour (nodes);
}
```

No array of connections is mentioned in this code, though you can see these connections at **figure 2**, and they definitely exist, as the rectangle can be moved by any border. When the array of nodes is passed as a parameter to the contour constructor and there is no array of connections, then such an array is organized automatically by connecting those nodes into an infinitive loop.

Move() and MoveContourPoint() methods in this case are absolutely identical to the previous one.

**Case 3**. *Rectangle*.

Moving by border; resizing by corners and middle points.

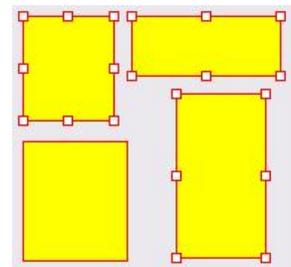

**Fig. 3**

In this case one of four possibilities of resizing is passed as a parameter on initialization. Contour always has eight nodes, but depending on this parameter of possible resizing:
- some nodes can be declared non-moveable, and thus are not shown;
- different number of nodes is shown (**figure 3**);
- the cursor shape above some of these nodes changes.

Four of the nodes are in the corners, other four – in the middle of the sides; indexing of the nodes starts in the left-top corner and goes clockwise. Even if some nodes are not shown as small squares, they still exist at the same places, only their area is null.

```
public override void DefineContour ()
{
    …
    ContourApex [] nodes = new ContourApex [8];
    switch (resize)
    {
        case ContourResize .Any:
            nodes = new ContourApex [8] {
                new ContourApex (0, pt [0], MovementFreedom.Any, Cursors.SizeNWSE),
                new ContourApex (1, pt [1], MovementFreedom.NS, Cursors.SizeNS),
                new ContourApex (2, pt [2], MovementFreedom.Any, Cursors.SizeNESW),
                new ContourApex (3, pt [3], MovementFreedom.WE, Cursors.SizeWE),
                new ContourApex (4, pt [4], MovementFreedom.Any, Cursors.SizeNWSE),
                new ContourApex (5, pt [5], MovementFreedom.NS, Cursors.SizeNS),
                new ContourApex (6, pt [6], MovementFreedom.Any, Cursors.SizeNESW),
                new ContourApex (7, pt [7], MovementFreedom.WE, Cursors.SizeWE)
            };
            break;
        case ContourResize .NS:
            for (int i = 0; i < nodes .Length; i++)
            {
                nodes [i] = new ContourApex (i, pt [i], MovementFreedom .NS,
                                                        Cursors .SizeNS);
            }
            nodes [3] = new ContourApex (3, pt [3], MovementFreedom .None,
```

```
                                            Cursors .SizeAll);
            nodes [7] = new ContourApex (7, pt [7], MovementFreedom .None,
                                            Cursors .SizeAll);
            break;
        …
        case ContourResize .None:
            for (int i = 0; i < nodes .Length; i++)
            {
                nodes [i] = new ContourApex (i, pt [i], MovementFreedom .None,
                                                Cursors .SizeAll);
            }
            break;
    }
    contour = new Contour (nodes);
}
```

`MoveContourPoint()` method is similar to case 1, only the number of nodes is greater here and an additional checking for the resizing, started by the corner nodes, is needed.

```
public override bool MoveContourPoint (int i, int cx, int cy, …, MouseButtons btn)
{
    bool bRet = false;
    MovementFreedom mf;
    if (btn == MouseButtons .Left)
    {
        if (i == 0)      // LT
        {
            contour .GetNodeFreedom (i, out mf);
            if (rc .Height - cy >= minH  &&
                    (mf == MovementFreedom.Any ||  mf == MovementFreedom.NS))
            {
                rc .Y += cy;
                rc .Height -= cy;
                bRet = true;
            }
            if (rc .Width - cx >= minW &&
                    (mf == MovementFreedom .Any || mf == MovementFreedom .WE))
            {
                rc .X += cx;
                rc .Width -= cx;
                bRet = true;
            }
        }
```

For special rectangles you may prefer to use contours of similar design, but with less not null nodes. For example, for a thin rectangle, which must be resized only horizontally, it maybe enough to have two square nodes only in the middle of the left and right sides. There are also other very interesting contours for rectangles; I'll write about them further on, but they can be used only with the graphical rectangular objects and not with the controls. Contours, similar to described here in case 3, are used with all the controls, involved in moving and resizing.

**Case 4**. *Graph*.

The proposed system of nodes and connections is ideal for organizing the moveable graph. It doesn't matter if all the nodes are connected to other parts of the graph, as on **figure 4**, or there are disjoint parts in the graph, or even some stand alone nodes; in any case, each node can be moved separately, and the whole graph can be moved by any connection.

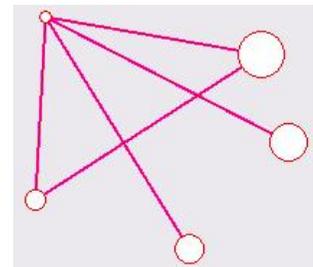

**Fig. 4**

```
public class CaseC_Graph : GraphicalObject
{
    Point [] pt;
    Color [] clr;
    int [] radius;
```

```
        List<GraphConnection> link = new List<GraphConnection> ();
```

In the previous cases, all nodes of a contour, which were not null, were identical, but there is no such requirement, and here each node has its unique size.

```
public override void DefineContour ()
{
    ContourApex [] nodes = new ContourApex [pt .Length];
    for (int i = 0; i < pt .Length; i++)
    {
        nodes [i] = new ContourApex (i, pt [i]);
        nodes [i] .SenseAreaForm = NodeForm.Circle;
        nodes [i] .SenseAreaSize = radius [i];
    }
    …
```

As I mentioned before, the code for individual movement of the nodes doesn't necessarily increase with the growth of the nodes' number. A graph might have any number of nodes, but the MoveContourPoint() method will be always the same.

```
public override bool MoveContourPoint (int i, int cx, int cy, …, MouseButtons btn)
{
    bool bRet = false;
    if (btn == MouseButtons .Left)
    {
        pt [i] .X += cx;
        pt [i] .Y += cy;
        bRet = true;
    }
    return (bRet);
}
```

Cases 1, 2, 3, and 4 represent the classical type of contour, in which relatively small and sparsely placed nodes are used for resizing or reconfiguring of objects and the lengthy enough connections between these nodes are used for objects' moving. Nodes can be also used for moving of the whole object; for such purpose they are usually enlarged to cover as much as possible of the object's area, and then moving is started in any (or nearly any) point of the area.

**Case 5**. *Regular polygon*.

Moving by a single circular node; no resizing.

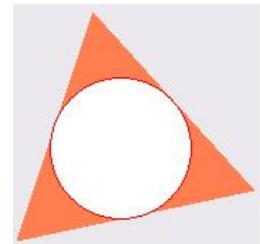

The proposed contour for moving (no resizing!) regular polygons comsists of a single circular node: this is a circle, inscribed into polygon. **Figure 5** demonstrates the worst possible case of moving such polygons, as only 60 percent of the inner area of a regular triangle is covered by such node, so the chances are really high that you'll press at the inner point and the triangle will not move. For the square, such node covers 79 percent of the inner area, and with the increasing number of vertices, the area, not covered with the circular node, diminishes, and such a contour looks not so bad. The definition of a contour, consisting of a single element, can't be too complicated.

**Fig. 5**

```
public override void DefineContour ()
{
    ContourApex [] ca = new ContourApex [1] { new ContourApex (0, ptC) };
    ca [0] .SenseAreaSize = nRadius;
    ca [0] .SenseAreaForm = NodeForm .Circle;
    contour = new Contour (ca, null);
}
```

What differs this case from all the previous, is the MoveContourPoint() method. Because the node here is used not for resizing, but for moving of an object, it can simply call the Move() method. Pay attention that this is the first of the samples, in which the MoveContourPoint() method doesn't use the node's index and simply ignores this parameter. This contour has only one node, further on the samples will have different number of nodes, but if all of them are used only for moving of an object, then the MoveContourPoint() method for such samples are often as simple as here.

```
public override bool MoveContourPoint (int i, int cx, int cy, …, MouseButtons btn)
{
    bool bRet = false;
    if (catcher == MouseButtons .Left)
    {
        Move (cx, cy);
        bRet = true;
    }
    return (bRet);
}
```

**Case 6**. *Rectangle*.

    Moving by any inner point; no resizing.

The enlarged nodes are often used for moving of the objects not by themselves, but in combination with connections. The only purpose of connections is to move objects; if you want to do it by any inner point of an object, you increase the sensitivity of connections to fill maximum of the area. But the sensitive area of any connection has a "sausage" shape, and for some objects of an arbitrary form the results maybe not satisfactory. In such cases an addition of nodes, used for moving of an object, is often a good solution. **Figure 6** demonstrates the combination of two enlarged square nodes and their connection; such system covers the whole area of any rectangle and allows it to be moved by any inner point.

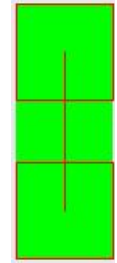

**Fig. 6**

```
public override void DefineContour ()
{
    int half = Math .Min (rc .Width, rc .Height) / 2;
    Point [] pt = new Point [2];
    pt [0] = new Point (rc .Left + half, rc .Top + half);
    if (rc .Width >= rc .Height) {
        pt [1] = new Point (rc .Right – 1 – half, pt [0] .Y);
    } else {
        pt [1] = new Point (pt [0] .X, rc .Bottom – 1 – half);
    }
    ContourApex [] ca = new ContourApex [2];
    for (int i = 0; i < 2; i++)
    {
        ca [i] = new ContourApex (i, pt [i], Cursors .SizeAll);
        ca [i] .SenseAreaSize = half;
        ca [i] .SenseAreaClearance = false;
    }
    contour = new Contour (ca);
    contour .ConnectionsSensitivity = half;
}
```

The size of the nodes is defined by the lesser dimension of the rectangle, so these nodes fill the ends of the rectangle from side to side. Depending on the rectangle size, these nodes can stay far away from each other (as on **figure 6**), overlap or even be positioned exactly at the same spot, if a rectangle is a square by itself. The overlapping of the nodes is not a problem, as all of them are used for moving of the object, and it doesn't matter by which node this movement is started. There are several important details in this contour:

- Sensitivity of the connection is enlarged to the sides of the rectangle.
- I took out the clearence of the nodes on visualization. If you keep it `true`, and the contour covers the whole area of an object, then you'll not see such object at all, when the contours are visualized.
- The cursor shape above the nodes is changed to `Cursors.SizeAll`. This is the standard shape of a cursor above any connection area; as nodes and connections in this case are used for the same purpose – moving of an object, I unified the cursor shape above all parts of the rectangle.

**Case 7**. *Rectangle*.

    Moving by any inner point; no resizing.

Nodes can be linked with each other by connections or can be used as stand alone objects; the number of such nodes, not connected to any other elements, is not limited. Contour can consist of a set of standing alone nodes (**figure 7**).

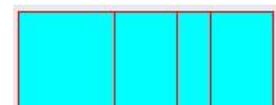

**Fig. 7**

```
public override void DefineContour ()
{
    int nodeSide;
    Point [] pt = Auxi_Geometry .ManySquaresOnRectangle (rc, out nodeSide);
    int half = nodeSide / 2;
    ContourApex [] ca = new ContourApex [pt .Length];
    for (int i = 0; i < pt .Length; i++)
    {
        ca [i] = new ContourApex (i, pt [i]);
        ca [i] .SenseAreaSize = half;
        ca [i] .SenseAreaClearance = false;
    }
    contour = new Contour (ca, null);
}
```

The design of contour is very similar to the previous case. The size of the nodes is defined by the lesser dimension of the rectangle; the nodes stand side by side to each other, only the last two nodes may overlap. The calculation of the nodes' centers is simple enough; only to shorten the code here I substitute it by a call to an auxiliary method, which returns the array of such points.

All the previous cases used two types of nodes: square and circle. There is one more type of nodes – *polygon nodes* – which are very useful in different situations. One of them, when there is a requirement to move an object by any inner point, but there is no easy and good enough way to cover the whole area only by the square and circular nodes.

**Case 8**. *Screw-nut*.

    Moving by any inner point; no resizing.

A screw-nut (**figure 8**) has six vertices on each border (inner and outer); contour consists of six trapeziums.

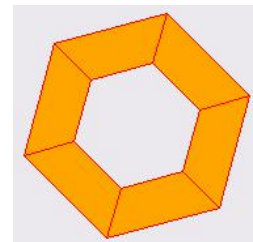

**Fig. 8**

```
public override void DefineContour ()
{
    Point [] ptIn = new Point [6];
    Point [] ptOut = new Point [6];
    double ang;
    for (int i = 0; i < 6; i++)
    {
        ang = angle + 2 * Math .PI * i / 6;
        ptIn [i] = new Point (ptC .X + (int) (rInner * Math .Cos (ang)),
                              ptC .Y - (int) (rInner * Math .Sin (ang)));
        ptOut [i] = new Point (ptC .X + (int) (rOuter * Math .Cos (ang)),
                               ptC .Y - (int) (rOuter * Math .Sin (ang)));
    }
    ContourApex [] ca = new ContourApex [6];
    int i1;
    for (int i = 0; i < 6; i++)
    {
        i1 = (i + 1) % 6;
        ca [i] = new ContourApex (i, new Point [] { ptIn [i], ptIn [i1],
                                                    ptOut [i1], ptOut [i] });
        ca [i] .SenseAreaClearance = false;
    }
    contour = new Contour (ca, null);
}
```

**Case 9**. *Rectangle*.

    Moving by any inner point; resizing by any border point.

Polygon nodes can be used instead of connections for moving of the objects; they can also play the standard nodes' role of resizing an object; in the proposed contour there are polygon nodes of both types (**figure 9**). Usually, when an object can be moved by any inner point and resized by any border point, there is no need in visualization of contours. But this is a sample for explanation of contour design, and the nodes on border are shown only for this purpose. As seen at this figure, there are areas, where from two to four nodes overlap, so the order of nodes in the contour is very important, as nodes are checked for possible movement according with their order. As I want the resizing to have higher priority than moving, and the resizing by corner to have the highest priority, then I use such order of nodes:

1. Four circular nodes in the corners.
2. Four polygon nodes on borders.
3. One polygon node inside.

```csharp
public override void DefineContour ()
{
    int half = 3;
    int cornerradius = 6;                   // 2 * half
    ContourApex [] nodes = new ContourApex [4 + 4 + 1];
    nodes [0] = new ContourApex (0, new Point (rc .Left, rc .Top),
                                    Cursors .SizeNWSE);
    nodes [1] = new ContourApex (1, new Point (rc .Right, rc .Top),
                                    Cursors .SizeNESW);
    nodes [2] = new ContourApex (2, new Point (rc .Right, rc .Bottom),
                                    Cursors .SizeNWSE);
    nodes [3] = new ContourApex (3, new Point (rc .Left, rc .Bottom),
                                    Cursors .SizeNESW);
    for (int i = 0; i <= 3; i++)
    {
        nodes [i] .SenseAreaForm = NodeForm .Circle;
        nodes [i] .SenseAreaSize = cornerradius;
    }
    nodes [4] = new ContourApex (4, new Point [] {
                                        new Point (rc.Left - half, rc.Top),
                                        new Point (rc.Left + half, rc.Top),
                                        new Point (rc.Left + half, rc.Bottom),
                                        new Point (rc.Left - half, rc.Bottom)},
                                MovementFreedom .WE, Cursors .SizeWE);
    …
    contour = new Contour (nodes, null);
}
```

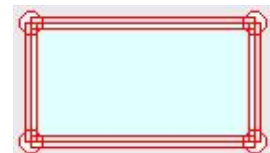

**Fig. 9**

Eight nodes in the corners and on borders are used for resizing, the biggest node, covering the whole rectangle, – for moving. The increased number of nodes resulted in the growth of the `MoveContourPoint()` method, but the code for different nodes duplicate each other, and all the parts of this method are really primitive.

All the previous cases used relatively small number of nodes. There are other cases, when the special technique of contour design must be used. These contours use the same type of nodes that were demonstrated in a lot of previous samples – square and circle nodes, but the amount of nodes can be huge. I named such contours *N-node contours*.

Contours are used for moving and resizing of objects. The most natural way for moving of the objects is to grab them by any inner point. Polygon nodes appeared as a good solution for the cases, when the area of an object cannot be covered in a satisfactory only by the square or circular nodes. The most natural way for resizing is to start it by any border point. In some situations it can be done by polygon nodes (case 9), but there are situations when such standard methods of contour design do not work well. In such cases the N-node contours is usually a very good solution.

N-node contours can consist of a huge number of small nodes, these nodes overlap with each other and cover the borders in such a way that they can be grabbed by any point. Usually the behaviour of all these nodes is identical, so the significant increase in the nodes' number doesn't increase the number of code lines in the `MoveContourPoint()` method.

When contour includes not one big set of identical nodes, but several sets of nodes with different characteristics, then there can be a situation with resizing, which requires the special method of dealing with the contour. Further on I'll write about the `Mover` class; an object of this class supervises the whole process of moving / resizing. The process starts, when `Mover` catches an object by one of the connections or nodes. Consider the situation, when resizing is started by grabbing one of the nodes. One of the most important parameters is the index of the grabbed node, as the `MoveContourPoint()` method defines the possible resizing by analyzing this index. When the size of an object changes, the contour may require different number of nodes to cover the changed border. In the redesigned contour the node with the same index (as was caught) may have different parameters, which will result in the change of the node's behaviour in the middle of the resizing process. To avoid this situation, the number of nodes in the contour must be fixed throughout the resizing process, and the contour must be redesigned only at the end of the resizing. I'll demonstrate this further on in one of the samples.

**Case 10**. *Circle*.

　　　Moving by any inner point; resizing by any border point.

A border of this circle (**figure 10**) is covered with a set of slightly overlapping small circular nodes, which are used for resizing; the whole inner area is covered by one enlarged circular node, which is used for moving. The number of the small nodes is defined by the length of the circumference. All the small nodes on the border are identical in behaviour and absolutely different from the circle in the middle. The big circle node receives index 0, the small nodes on the border will have indices from 1 and up, and then it doesn't matter, how the number of circles on the border changes throughout the process of resizing, as the MoveContourPoint() method will distinguish the nodes' behaviour between the index 0 and all others.

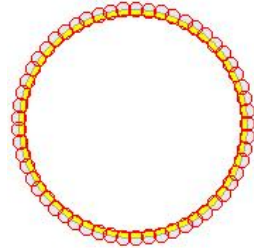

**Fig. 10**

```
public override void DefineContour ()
{
    int nOnPerimeter = Convert .ToInt32 ((2*Math.PI * nrBig) / distanceNeighbours);
    ContourApex [] nodes = new ContourApex [nOnPerimeter + 1];
    nodes [0] = new ContourApex (0, ptCenter, Cursors .SizeAll);
    nodes [0] .SenseAreaForm = NodeForm .Circle;
    nodes [0] .SenseAreaSize = nrBig – nrSmall + 1;
    for (int i = 1; i <= nOnPerimeter; i++)
    {
        nodes [i] = new ContourApex (i, Auxi_Geometry .PointToPoint (ptCenter,
                                     2 * Math .PI * (i - 1) / nOnPerimeter, nrBig));
        nodes [i] .SenseAreaForm = NodeForm .Circle;
        nodes [i] .SenseAreaSize = nrSmall;
        nodes [i] .SenseAreaClearance = false;
    }
    contour = new Contour (nodes, null);
}
```

The individual movement of the nodes play different role, depending on the index of the node: the big one (with zero index) moves the whole circle, any other node changes the size of the circle.

```
public override bool MoveContourPoint (int i, int cx, int cy, …, MouseButtons btn)
{
    bool bRet = false;
    if (catcher == MouseButtons .Left)
    {
        if (i == 0)
        {
            Move (cx, cy);
        }
        else
        {
            int nRadNew = Convert.ToInt32 (Auxi_Geometry.Distance (ptCenter, ptM));
            if (nRadNew != nrBig && nRadNew >= nMinRadius) {
                nrBig = nRadNew;
                bRet = true;
            }
        }
    }
    return (bRet);
}
```

During the whole process of resizing, the small nodes at any moment cover the border without any gaps; this means that the contour is cosntantly redesigned.

**Case 11**. *Ring*.

Moving by any inner point; resizing by any border point.

The contour for a ring on **figure 11** combines the ideas of N-node contour and polygon nodes. Both circular borders – inner and outer – are covered by a set of small overlapping nodes; these nodes are used for resizing in the same way, as was done with the circle in case 10. The inner area of a ring is covered with a set of polygon (trapezoid) nodes; these nodes are used for moving, so the whole ring can be moved by any inner point.

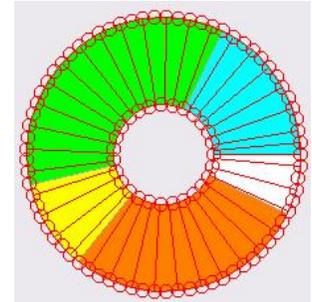

**Fig. 11**

```
public override void DefineContour ()
{
    ContourApex [] nodes = new ContourApex [nNodes];
    Point [] ptOuter = new Point [nNodesOnOuter];
    for (int i = 0; i < nNodesOnOuter; i++)
    {
        ptOuter [i] = Auxi_Geometry .PointToPoint (ptCenter,
                                   2 * Math .PI * i / nNodesOnOuter, nrOuter);
    }
    …
    for (int i = 0; i < nNodesPoly; i++)
    {
        int jOutPt = (i * 2) % nNodesOnOuter;
        p0_out = ptOuter [jOutPt];
        p1_out = ptOuter [(jOutPt + 2) % nNodesOnOuter];
        p0_in = Auxi_Geometry .PointOnLine (ptCenter, p0_out, ratioRadiuses);
        p1_in = Auxi_Geometry .PointOnLine (ptCenter, p1_out, ratioRadiuses);
        nodes [nSmall + i] = new ContourApex (nSmall + i,
                                   new Point [] { p0_in, p0_out, p1_out, p1_in });
    }
    contour = new Contour (nodes, null);
}
```

MoveContourPoint() method only slightly differes from the previous case as not all the nodes on the borders have identical behaviour, but they are divided into two sets, and the nodes from these sets are changed in different way for the possibility of resizing.

As shown on **figure 11**, the nodes on he borders only slightly overlap; just enough to make a ring resizable by any border point. If you try to enlarge radius of any border, the small circles on this border will move apart, showing the gaps between them; if you shrink the border, the nodes will overlap more and more. When you release the button, the picture changes to normal. How is it organized?

When the ring (of the NRing class) is constructed, the needed number of the nodes on each border and the needed number of the trapezoid nodes are defined, based on two radiuses, and the contour is designed. Throughout the resizing, these numbers are not changed, the Mover is constantly working with the node that was caught at the beginning, and the whole process goes smoothly. When an object is released, Mover checks the type of this object, and if it belongs to the NRing class, then the method of redefining contour for this object is called.

```
private void OnMouseUp (object sender, MouseEventArgs e)
{
    if (mover .Release ())
    {
        int iCaughtObject = mover .WasCaughtObject;
        if (mover [iCaughtObject] .SourceGraphical is NRing)
        {
            (mover [iCaughtObject] .SourceGraphical as NRing) .RedefineContour ();
            Invalidate ();
        }
    }
}
```

NRing.RedefineContour() method is very simple: it only defines the needed number of nodes on the borders according with the current radiuses and calls the NRing.DefineContour() method.

```
public void RedefineContour ()
{
    nNodesOnOuter = Convert .ToInt32 ((2 * Math.PI * nrOuter) / distanceNeighbours);
    nNodesOnInner = Convert .ToInt32 ((2 * Math.PI * nrInner) / distanceNeighbours);
    nNodesPoly = nNodesOnOuter / 2 + 1;
    DefineContour ();
}
```

All the previous cases were about the graphical objects; now let's look at the controls. Though I want the controls to be involved in the moving / resizing process exactly in the same way as graphical objects, but their special features put some limitations and require some different steps in design.

**Case 12**. *Controls*.
Moving by borders; resizing by the special border areas – nodes on the borders.

Users don't need to know the difference between controls and graphical objects, and usually they don't even know these words. But all users know that:

- There are buttons, which they can click with a mouse to start some process.
- There are check boxes and radio buttons, which they can select or unselect to inform the program about their choices.
- There are lists of strings, in which they can select one or several needed items.

What is important for turning all controls into easily moveable / resizable, is the fact that inner areas of all the controls are already used for well known and expected mouse events and reactions, so these areas cannot be used for contours design. The only way to catch a control and move it or resize it, is to grab it by the border! On one side, it limits the design of the contours for controls. On the other side, as all the standard controls have rectangular shape, this makes the moving / resizing of controls absolutely simple for users, as all the controls will be moved / resized in exactly identical way. This also simplifies the process of contour design: the contour is defined by `Mover`, when the control is included into its List<>; some variations in this contour depend on the set of parameters.

```
mover .Insert (0, buttonContours);
mover .Insert (0, listInfo, ContourResize .Any, 250, 500, 80, 240);
```

The first line makes the button moveable, but non-resizable; the second line makes the ListView moveable and defines the ranges, in which the sizes of this control can change. **Figure 12** demonstrates the contour of this control.

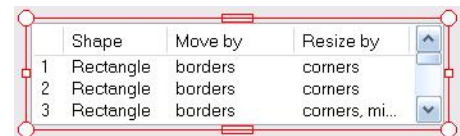

**Fig. 12**

On the upper level, windows are moved by the title bar and resized by any border point on other three sides. The same solution can't be used for controls, as they have no title bars. The sensitive areas for moving and resizing must be placed somewhere close to the controls' borders. There is a well known (and thus expected) solution for this problem: a rectangle is resized via the small areas in the corners and in the middle of the sides, and moved by all other parts of the border. This is done in several applications, but all of them visualize these sensitive areas for resizing. As my moving / resizing works regardless of whether the contours are shown or not, I use the same type of contour, but a bit modified.

- The corner nodes are slightly enlarged.
- All four corner nodes are in use even when the control can be resized only in one direction (horizontally or vertically).
- The nodes in the middle become larger, when the size of the control increases.

Without these small changes, the controls have the same four variants of contour that are shown for rectangle in case 3.

One more remark about this contour. Cases 8, 9, and 11 already used the polygon nodes, but this sample is the first sample of such node, for which the position of the related real point is important, because this is the only sample with polygon nodes, connected to other nodes, and the connection goes to some real point.

These samples demonstrate the design of the different contours. The contours of very complicated objects can use exactly the same contours or be based on these ideas.

**Contour summary**

- A contour can consist of any number of nodes and their connections.
- The minimum number of nodes is 1. A contour may consist of a single node without any connections.
- A contour may consist of a set of disjoint parts; each part may consist of arbitrary connected nodes or it can be a single node.
- Connections may exist only between nodes; there must be a node on both ends of any connection.
- Nodes can be moved individually thus allowing you to reconfigure the object.
- By grabbing any connection, the whole object can be moved.
- Node can be enlarged to cover as much of the objects area as possible. Such enlarged nodes are often used instead of connections for moving of the whole object.
- A contour may consist of a series of connections between empty nodes. This makes an object moveable, but not resizable.
- Each of the nodes has its own parameters. By connecting the nodes with different types of allowed individual movement, it is easy to allow resizing along one direction but prohibit it along another. This means organizing a limited reconfiguration.
- It doesn't matter that some contours may represent graphical objects and other contours are used for controls or groups of controls ([3, 4]). All contours are treated in the same way, thus allowing the user to change easily the inner view of any application.

## Moving and resizing

Contour is an additional special feature of any object that allows it to be turned into moveable / resizable. But the moving / resizing process have to be organized in some way; I prefer to do it with a mouse and only with a mouse, thus this process is organized via the three mouse events: `MouseDown`, `MouseMove` and `MouseUp`. And this process is supervised by a specially designed class `Mover`.

An object of the `Mover` class plays the same role, as the Windows system plays to the windows. Any object inside an application can be involved in moving / resizing process if:

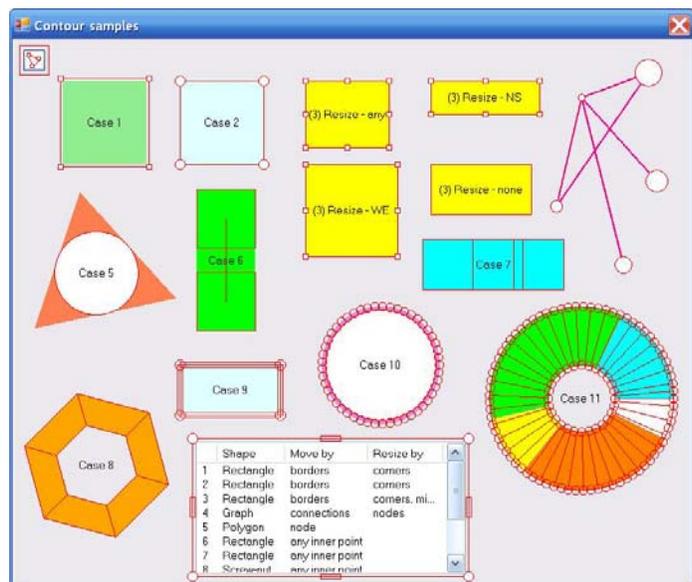

1. This object's class is derived from the `GraphicalObject` and has an implementation of three methods: `DefineContour()`, `Move()`, and `MoveContourPoint()`.
2. Object itself is registered with the `Mover`.

The first thing only gives an object an ability to be moveable / resizable. The second thing really makes an object moveable / resizable, as `Mover` organizes this process only for the registered objects.

`Mover` doesn't know anything about the real objects; it deals only with the contours of the registered objects. Each `Mover` has its personal list of registered objects; there are no restrictions on the number of `Mover` objects in the form and on the number or type of the 

**Fig.13** Different types of contours (**Form_ContourSamples.cs** from the **Test_MoveGraphLibrary** application)

objects, registered with each of them. The need for several `Mover` objects in the form is a very rare thing; I described this in [2, 3, 4], but in general one `Mover` is enough.

**Figure 13** demonstrates the view of the **Form_ContourSamples.cs** from the **Test_MoveGraphLibrary** application. This form includes not only all the previously mentioned samples, but also a `ListView` with information about all of them. Let's see, how the moving / resizing process is organized for such a collection of different objects with absolutely different contours.

　　　　`Mover mover;`

Supervising the moving / resizing process means not only the pure moving and resizing of objects, but also:

- preventing an accidental moving the objects out of view;
- informing about the possibility of moving / resizing by changing the mouse cursor;
- giving the detailed information about the currently moved / resized object, or about the object that is currently under the cursor;
- drawing the contours, if needed.

When an application consists of only standard elements fixed at design time, then there are no problems with some elements going out of view.  When an application is based on the moveable elements, this thing can happen easily.  It can happen accidentally, or it can be done on purpose by user, because this is a part of forms' customization that moveable elements allow to organize.  If an element is moved across upper or left border and dropped there, then there is no way to recover such an element; if it is dropped across the right or bottom border, it can be recovered by enlarging the form.  `Mover` can control this process and allows to set three different levels of moving elements across the borders [3, 4].

Registering of the elements with the `Mover` is very simple.

```
mover .Add (rc1);
mover .Insert (0, rc2);
```

For graphical objects, the possible restrictions on moving / resizing are implemented in the `MoveContourPoint()` method.  Controls have no such methods; the restrictions on their resizing can be defined by a set of parameters at the moment of the registration.

```
mover .Insert (0, listInfo, ContourResize .Any, 250, 500, 80, 240);
```

As I mentioned before, the three mouse events are used for moving / resizing; usually the code for them is extremely simple.  If the ring (class `NRing`, case 11) was absent in the above mentioned form, then the code for those mouse events would look like this.

```
private void OnMouseDown (object sender, MouseEventArgs e)
{
    mover .Catch (e .Location);
}
private void OnMouseUp (object sender, MouseEventArgs e)
{
    mover .Release ();
}
private void OnMouseMove (object sender, MouseEventArgs e)
{
    if (mover .Move (e .Location))
    {
        Invalidate ();
    }
}
```

I have already explained, why this ring needs some modification of the `OnMouseUp()` method.  Each class, which has a contour that can't be changed throughout the resizing but needs this to be done at the end of it, will require a couple of additional lines in this method.

```
private void OnMouseUp (object sender, MouseEventArgs e)
{
    if (mover .Release ())
    {
        int iCaughtObject = mover .WasCaughtObject;
        if (mover [iCaughtObject] .SourceGraphical is NRing)
        {
            (mover [iCaughtObject] .SourceGraphical as NRing) .RedefineContour ();
            Invalidate ();
        }
    }
}
```

Some of the elements in application may need rotation.  There is no difference for `Mover` to supervise either the forward movement, or rotation.  The whole process is described in details in [4]; usually it would require couple of additional lines in the `OnMouseDown()` method.

**Moving / resizing summary**

- To make any graphical object moveable / resizable, it must be derived from `GraphicalObject`.
- Controls can be included into moving / resizing as they are.
- Any combination of elements can organize a set of synchronously moving objects.
- To organize moving / resizing process, there must be an object of `Mover` class.

    ```
    Mover mover;
    ```

- To prevent the accidental moving of elements out of view, the parent form must be mentioned during the `Mover`'s initialization

    ```
    mover = new Mover (this);
    ```

- `Mover` will supervise the whole moving / resizing process, but only for graphical objects and controls that are included into its List<>.

    ```
    mover .Add (…);
    mover .Insert (…);
    ```

- Moving and resizing are done with a mouse, and the whole process is organized via the three standard mouse events: `MouseDown`, `MouseUp` and `MouseMove`.
- `MouseDown` starts moving / resizing by grabbing either the whole object (moving), or a single node (resizing). The only mandatory line of code in this method is

    ```
    mover .Catch (…);
    ```

- `MouseUp` ends moving / resizing by releasing any object that could be involved in the process. The only mandatory line of code in this method is

    ```
    mover .Release ();
    ```

- `MouseMove` moves the whole object or a single node. There is one mandatory line of code in this method, but in order to see the movement, the `Paint` method must be called

    ```
    if (mover .Move (mea .Location))
    {
        Invalidate ();
    }
    ```

- If contours must be shown, then one of the available drawing methods must be used, for example,

    ```
    mover .DrawContours (grfx);
    ```

- If needed, several `Mover` objects can be used to organize the whole moving / resizing process. Each `Mover` deals only with the objects from its own List<>.

## Conclusion

These are the basic ideas in contour design and turning the screen objects into moveable / resizable. The consequences of using such elements – the design of applications exclusively on resizable and moveable objects – is a revolutionary thing in applications design. A lot of interesting results and more detailed explanation can be seen at (and downloaded from) www.sourceforge.net in the project **MoveableGraphics**.

Dr. Sergey Andreyev ( andreyev_sergey@yahoo.com )

September 2008